# Solution to the measurement problem within linear, unitary, no collapse, no particle quantum mechanics.

Casey Blood
Rutgers University, retired
CaseyBlood@gmail.com

The measurement problem is to explain why a system which is in a linear combination of states appears, upon measurement, to be in just one of those states. The solution given here is to first show that if one assumes linear, unitary, no collapse, no particle quantum mechanics, the different versions of reality are completely isolated from each other. This then implies only the eigenstates appropriate to the measurement setup will be perceived. In a Stern-Gerlach experiment, for example, only spin up or spin down will be perceived, but never a combination of the two.

**1. Introduction.**

Quantum mechanics is a highly successful theory in that it correctly predicts the quantitative properties of matter from the subatomic to the astronomical levels. But it appears on the surface at least, to fail qualitative. If Schrodinger's cat is simultaneously dead and alive in the wave function, for example, what prevents us from perceiving the cat both dead and alive at the same time? In discussing this problem [1-3], it is often assumed that because we perceive only one outcome in a measurement, the wave function has collapsed to one outcome. Or else it is assumed there is, after measurement, an actual object, different from the wave function, which has the characteristics of just one of the outcomes.

We take a different tack. We suppose instead that physical existence always conforms to the mathematics and concepts of basic quantum mechanics in which (1) only the state vectors exist (no particles), (2) there is no collapse, and (3) the time translation operator is linear and unitary. That is, we assume Everett's many-worlds [4] point of view, but without his treatment of probability. Most physicists assume the measurement problem cannot be solved within basic quantum mechanics. But, because all the *quantitative* results (except probability) follow from the mathematics of basic quantum mechanics, it would be surprising if it failed to properly describe the *qualitative* aspects of measurements.

And indeed we find there is no surprise; basic quantum mechanics leads only to those perceptions that are seen in measurements in the laboratory. The mathematics shows that a combination of cat dead and cat alive, or spin up and spin down, will never be seen. Our argument is similar to Everett's but the methods used in the next two sections fill in gaps in his reasoning.

**2. Isolation of versions of reality.**

To see how the measurement problem is solved within basic quantum mechanics, we first need to establish a fundamental property of the state vectors. Using a spin ½ Stern-Gerlach experiment as an example, we show that each version of reality evolves in time as if the other versions were not there.

After the Stern-Gerlach magnet, the $|-1/2\rangle$ version of the spin ½ particle (particle-like wave function) travels on path 1 and the $|+1/2\rangle$ version travels on path 2. Detector $|\text{Det}-\rangle$ is on path 1 and $|\text{Det}+\rangle$ on path 2. The state vector of the particle-like wave function at time 0, after the magnet but before detection, is a linear combination

$$|\psi(0)\rangle = a(1)|-1/2\rangle + a(2)|+1/2\rangle = a(1)|-\rangle + a(2)|+\rangle \tag{1}$$

while the state of the whole system, including an observer who looks at the readings on the detector dials, is

$$\begin{aligned}|\Psi(0)\rangle &= a(1)|\text{version 1}\rangle + a(2)|\text{version 2}\rangle \\ |\text{version 1}\rangle &= |-\rangle|\text{Det}-, \text{no}\rangle|\text{Det}+, \text{no}\rangle|\text{Obs sees no, no}\rangle \\ |\text{version 2}\rangle &= |+\rangle|\text{Det}-, \text{no}\rangle|\text{Det}+, \text{no}\rangle|\text{Obs sees no, no}\rangle\end{aligned} \tag{2}$$

with 'no' meaning no detection. The linear, unitary time translation operator U(t) takes the state to time t;

$$\begin{aligned}|\Psi(t)\rangle &= U(t)|\Psi(0)\rangle \\ &= a(1)U(t)[|\text{version 1}\rangle] + a(2)U(t)[|\text{version 2}\rangle]\end{aligned} \tag{3}$$

Now because U(t) is linear, U(t)[|version 1⟩] is independent of a(1) and a(2). Thus it can be calculated at a(1)=1, a(2)=0 (and then it will be the same no matter what the values of a(1) and a(2)). But when a(2)=0, |version 2⟩ is just not there. This means the time evolution of |version 1⟩ cannot be affected by anything that happens on the |version 2⟩ branch of reality. For example, if a photon is emitted by a version of the detector on branch 2, it will never be perceived by a version of the observer on branch 1. Because each evolves as if the other were not there, the two branches cannot interact in any way. Thus the branches constitute separate, completely isolated, non-communicating universes. Among other consequences, this implies different versions of the observer are never aware of each other.

### 3. Solution to the measurement problem.

The measurement problem is to explain why, after measurement, we always perceive results corresponding to either the single-version state |−⟩ or the single-version state |+⟩, but never to some linear combination, even though the initial wave function is a linear combination of the two states. It is often assumed there is no such explanation within basic quantum mechanics, so some modification such as collapse or the existence of actual particles is needed to account for the perception of only non-combination states. But that is not true.

To show this, we start from Eq. (3) and calculate $U(t)[|\text{version 1}\rangle]$ when a(1)=1, a(2)=0 (i.e., version 2 is simply not there). We know that as time progresses, the 'particle'-detector Hamiltonian will cause the |−⟩ state to change the |Det−⟩ state from |Det−, no⟩ to |Det−, yes⟩, but the |Det+⟩ state will stay at |Det+, no⟩. The dials on the versions of the detectors on branch 1 will emit photons that are perceived by the version of the observer on branch 1. But because of the isolation of the branches, version 1 of the observer will never perceive photons emitted by the dials of the version of the detectors on branch 2. (Note: It is assumed that the no and yes states of the detectors are orthogonal.) And similarly for branch 2. Thus we have, with the observer writing down what she perceives,

$$\begin{aligned}U(t)|\text{version 1}\rangle &= |-\rangle|\text{Det}-, \text{yes}\rangle|\text{Det}+, \text{no}\rangle|\text{ver. 1 of obs writes "I see only yes, no"}\rangle \\ U(t)|\text{version 2}\rangle &= |+\rangle|\text{Det}-, \text{no}\rangle|\text{Det}+, \text{yes}\rangle|\text{ver. 2 of obs writes "I see only no, yes"}\rangle\end{aligned} \tag{4}$$

When we put these back into Eq. (3), we get

$$|\Psi(t)\rangle = a(1)|-\rangle|\text{Det}-,\text{yes}\rangle|\text{Det}+,\text{no}\rangle|\text{ver. 1 of obs writes "I see only yes, no"}\rangle$$
$$+a(2)|+\rangle|\text{Det}-,\text{no}\rangle|\text{Det}+,\text{yes}\rangle|\text{ver. 2 of obs writes "I see only no, yes"}\rangle \quad (5)$$

Note that no basis (and in particular no 'preferred' basis) was used in arriving at these results. The isolation of branches, the particle-detector Hamiltonian, and the Hamiltonian for the detector-observer messenger photons absolutely determine the results of Eq. (5), independent of any basis. But suppose we decided to re-express this state in terms of a linear combination basis for the observer states after perception. Then, to agree with Eq. (5) and with the implications of the interaction Hamiltonians, the basis vectors will of necessity be linear combinations of the two observer states in Eq. (5), *including the writing*. Thus, even in a linear combination basis, only "I see only yes,no" or "I see only no,yes" will be written; there will never be a version of the observer who writes "I see something besides a single, classical version of reality."

Note also that the only observing entities in basic quantum mechanics are the quantum mechanical versions of the observers. There is, for example, no over-arching Observer who is simultaneously aware of what both version 1 of the observer and version 2 of the observer perceive.

If basic quantum mechanics is to fail in its description of the possible outcomes of measurements then it must lead to the perception of linear combinations which are never seen in the laboratory. We call such perceptions non-classical, where 'classical' means consistent with our everyday, classical-physics-like perceptions (and 'non-classical,' of course, means not consistent). The logic of the argument, then, is that if one claims basic quantum mechanics fails to properly describe measurements, one must be able to give a specific example where some version of the observer writes "I see something besides a single, classical version of reality." But we see from Eq. (5) and the argument about basis vectors that no such example can be constructed.

We conclude that for all practical purposes (that is, for all communicable perceptions), there will never be perception of anything besides classical versions of reality in basic quantum mechanics; the versions of the observer perceive either $|-\rangle$ or $|+\rangle$, never a linear combination. Thus there is no measurement 'problem' in basic quantum mechanics because its predictions do not conflict with our perception of only classical outcomes to measurements.

### 4. More complex measurement situations.

The same line of reasoning—using only basic quantum mechanics and the properties of the various Hamiltonians—can be applied to other measurements. Suppose we consider an electron wave function scattering off a proton wave function located in the center of a sphere covered with N grains of film. There will be N separate, grain-electron Hamiltonians, each of which will lead to a state, after scattering, with one and only one grain exposed. The logic of the previous section then shows that only one of the N versions of reality will be perceived. Thus the observer will perceive one and only one localized grain exposed, exactly *as if* there were a particulate electron that hit just one grain—even though we assume only wave functions exist!

The same reasoning, used repeatedly, can also account for the particle-like trajectories seen in bubble chambers [5].

Essentially the same analysis can be applied to the double slit experiment. There is, however, a complication here; the equivalents of the $|-\rangle$ and $|+\rangle$ states in that experiment interfere before detection. This requires a modification of the analysis, but the conclusion is the same; basic quantum mechanics leads to the perception of one and only one localized grain exposed per run. When taken together, these results imply that basic quantum mechanics can account for the particle-like property of localization.

This basic quantum mechanics approach, with no added interaction at a distance, also correctly describes the results perceived in experiments on entangled states—the Bell-Aspect experiment [6,7], the quantum eraser [8], the Ionicioiu-Terno version of the Wheeler delayed choice experiment [9] and so on.

We also note that the basic quantum mechanics approach has no difficulty in handling conditional results. If we have several successive beam splittings, for example, and if the results of early splittings are known, then basic quantum mechanics correctly predicts the successive possible perceptions.

There is one other point worth mentioning because it shows the wide range of the implications of basic quantum mechanics. The linear equations of basic quantum mechanics are invariant under the inhomogeneous Lorentz group and internal symmetry groups. This implies, through group representation theory, that the particle-like properties of mass, energy, momentum, spin, and charge are all properties of the state vectors (because the state vectors correspond to solutions of invariant equations). Note that this is a *derivation*(!) of the fact that matter, presumed to be made up of wave functions/state vectors alone, will possess the particle-like properties of mass, energy, momentum, spin (properly quantized), and charge (also properly quantized [10]), and that the appropriate conservation laws will apply to these quantities.

### 5. Agreement among observers.

To match our perceptions, it is not only necessary that each observer perceive a classical outcome; it is also necessary that observers agree on the outcome. But that automatically occurs in basic quantum mechanics because each version of reality, each branch, constitutes a closed system. So all perceptions are within a branch. To illustrate this in equation form, we give the two-observer analog of Eq. (5), asking the second observer to write whether she agrees with the first. The state vector is

$$\begin{aligned}|\Psi(t)\rangle = &\, a(1)|-\rangle|\text{Det}-,\text{yes}\rangle|\text{Det}+,\text{no}\rangle \\ &\quad |\text{ver. 1 of obs 1 writes "I see only yes, no"}\rangle \\ &\quad |\text{ver. 1 of obs 2 writes "I see only yes, no and I agree with obs 1"}\rangle \\ &\quad + \\ &\, a(2)|+\rangle|\text{Det}-,\text{no}\rangle|\text{Det}+,\text{yes}\rangle \\ &\quad |\text{ver. 2 of obs 1 writes "I see only no, yes"}\rangle \\ &\quad |\text{ver. 2 of obs 2 writes "I see only no, yes and I agree with obs 1"}\rangle\end{aligned} \quad (6)$$

The agreement between versions of two observers on the same branch occurs because both can perceive only the versions of the detectors and the other observer on that branch.

## 6. Limit to the success of basic quantum mechanics.

In spite of all its successes, basic quantum mechanics cannot be the whole story because, by itself, it does not lead to the probability law. To see this, suppose there are N separate outcomes of an experiment. Then after a single run, there are N isolated versions of reality. All versions are equally valid in the sense that no version of either the outcome or the observer is singled out in basic quantum mechanics as the only 'real' one. And there is no 'super observer' whose perceptions correspond to just one version of the observer. Thus, because there is no singling out of any sort, there can be no *probability* of perceiving a given version of reality. Or put more simply, the probability of observer version $i$ perceiving outcome $j$ is $\delta_{ij}$, regardless of the coefficients.

It may be that solving the probability problem will make our solution to the measurement problem beside the point. That would be true, for example, if evidence is found that collapse occurs (there is currently no such evidence). But our argument shows that no modification of basic quantum mechanics is necessary for solving the measurement problem.